\documentclass{moriond}

\usepackage{amsmath,amssymb,mathrsfs}

\def\be{\begin{equation}}
\def\ee{\end{equation}}
\def\bea{\begin{eqnarray}}
\def\eea{\end{eqnarray}}

\newcommand{\Photo}{}

\begin{document}
\vspace*{4cm}
\title{AXIONS AND AXION-LIKE PARTICLES}

\author{ A. RINGWALD }

\address{Deutsches Elektronen-Synchrotron DESY, Notkestr. 85,\\
D-22607 Hamburg, Germany}

\maketitle\abstracts{
The physics case for axions and axion-like particles is reviewed and an 
overview of ongoing and near-future laboratory searches  is
presented.}

\section{Introduction}

The Standard Model (SM) of particle physics describes the properties of known matter 
and forces to a fantastic accuracy. However, it can 
not be considered to be a complete 
and fundamental theory. Most notably, it does not give a satisfactory explanation for 
the values of its many parameters, it does not provide a predictive quantum theory of gravity, and it 
does not explain the origin of the dark sector of the Universe. In fact, the observation that 
nearly thirty percent of the energy content of the universe consists of dark matter 
provides a very strong case for the existence of particles beyond the SM. 

Intriguingly, dark matter candidates occur automatically in  a number of SM extensions which were 
originally motivated by completely other reasons. Prominent examples are the lightest supersymmetric (SUSY) 
partners, e.g. neutralinos (partners of the SM gauge or Higgs bosons) or gravitinos (partners of the 
graviton), in R-parity conserving 
SUSY extensions of the SM -- the latter being motivated by grand unification and the solution of the hierarchy problem. 
Another widely-discussed example are the axion -- occuring in extensions of the SM solving the strong
CP problem -- and axion-like particles (ALPs), which are often predicted by embeddings of the SM in string theory. 

In this contribution we review the theoretical motivation for the axion and ALPs (Sec.  \ref{sec:theory}), 
summarize the various hints for their existence from astrophysics and cosmology (Sec.  \ref{sec:hints}), 
and report on ongoing and near-future laboratory experiments and observatories (Sec.  \ref{sec:experiments}).

\section{\label{sec:theory}Theoretically Favored ALP Candidates}

Many extensions of the SM feature one or several  spontaneously broken global $U(1)_i$ symmetries, 
$i=1,\ldots, n_{\rm ax}$. At energies below their symmetry breaking scales $v_i$, Nambu-Goldstone boson fields $a_i^\prime$ arise as massless excitations of the angular 
part of the SM singlet complex scalar fields $\phi_i$ whose vacuum expectation values (vev) $\langle \phi_i\rangle =v_i/\sqrt{2}$ break the $U(1)_i$ symmetries: 
$\phi_i (x)= (v_i + \sigma_i(x)) {\rm e}^{i a_i^\prime (x)/v_i}/\sqrt{2}$.  Their interactions with SM particles, 
e.g. with gluons (described by the gluonic field strength $G_{\mu\nu}$), photons (described by the electromagnetic field strength $F_{\mu\nu}$), and electrons (described by the spinor $e$),  
are suppressed by inverse powers of the supposedly large symmetry breaking scales, $f_{a_i^\prime}=v_i\gg v$,
where $v=246$ GeV is the electroweak Higgs vev, 
\begin{eqnarray}
\label{ALP_leff}
\mathcal L & = &\frac{1}{2}\, \partial_\mu a_i^\prime\, \partial^\mu a_i^\prime
- \frac{\alpha_s}{8\pi} \left(\sum_{i=1}^{n_{\rm ax}} C^\prime_{ig} \frac{a_i^\prime}{f_{a_i^\prime}}\right)  G_{\mu\nu}^b \tilde{G}^{b,\mu\nu}
\\ 
\nonumber
&& - \frac{\alpha}{8\pi}
\left(\sum_{i=1}^{n_{\rm ax}}C^\prime_{i\gamma} \frac{a_i^\prime}{f_{a_i^\prime}}\right) \, F_{\mu\nu} \tilde{F}^{\mu\nu}
+ \frac{1}{2} \,
\left( \sum_{i=1}^{n_{\rm ax}}C^\prime_{ie}\frac{\partial_\mu a_i^\prime}{f_{a_i^\prime}} \,\right)\,\bar{e} \gamma^\mu\gamma_5 e
+\ldots \,.
\end{eqnarray}
Here, the couplings to the gluons, $C^\prime_{ig}$, and to the photons, $C^\prime_{i\gamma}$, arise from integrating out fermions with chirally anomalous $U(1)_i$ charge assignments.  
Particularly well-motivated examples for such Nambu-Goldstone bosons are: 
\begin{itemize}
\item The {\it axion} $A$ -- the particle excitation of the superposition of all Nambu-Goldstone fields $a_i^\prime$ coupling to the topological charge density in QCD, $q\equiv \frac{\alpha_s}{8\pi}   G_{\mu\nu}^b \tilde{G}^{b,\mu\nu}$, in Eq. (\ref{ALP_leff}), 
\begin{equation}
\frac{A}{f_A}\equiv\sum_{i=1}^{n_{\rm ax}} C^\prime_{ig} \frac{a_i^\prime}{f_{a_i^\prime}}. 
\label{axion_def}
\end{equation}
This field replaces the theta parameter in QCD by a dynamical quantity, $\theta_A (x)= A(x)/f_A$, spontaneously relaxing to zero, $\langle \theta_A\rangle =0$ -- thereby explaining the 
non-observation of strong CP violation\cite{Peccei:1977hh,Weinberg:1977ma,Wilczek:1977pj}. 
In fact, topological non-trivial gluonic fluctuations result in an  effective potential for $\theta_A$, 
\begin{equation}
V(\theta_A)=
\frac{ \chi (0)}{2} 
\theta_A^2 + {\mathcal O}(\theta_A^4)
\simeq 
\frac{ m_\pi^2 f_\pi^2}{2} 
\frac{m_u m_d}{(m_u+m_d)^2} 
\theta_A^2 + {\mathcal O}(\theta_A^4)
, \label{effpot}
\end{equation}
which has a localized minimum at $\theta_A=0$. Here, $\chi (0)=\langle Q^2\rangle\mid_{\theta =0}/\int d^4x$, with $Q=\int d^4x\, q(x)$, is the topological susceptibility,  
$m_\pi$ and $f_\pi$ are the mass and the decay constant of the pion, and $m_u$ and $m_d$ are the quark masses, 
respectively. Moreover, the topological fluctuations give the axion a small mass, which 
can be read off from the quadratic part in Eq.  (\ref{effpot}),   
\begin{equation}
m_A = \frac{m_\pi f_\pi}{f_A} \frac{\sqrt{m_u m_d}}{m_u+m_d}\simeq { 0.6\,  {\rm meV}}
         \times
         \left(
         \frac{10^{10}\, {\rm GeV}}{f_A}\right),
\label{axion_mass}
\end{equation}
rendering the axion a pseudo-Nambu-Goldstone boson. Due to the mixing with the neutral pion, the axion has 
a universal coupling to photons,
\begin{equation}
\label{photon_coupling_universal}
\mathcal L \supset
- \frac{g_{A\gamma}}{4}\,A\, F_{\mu\nu} \tilde{F}^{\mu\nu}
; \hspace{6ex}
        | g_{A\gamma} |
\sim \frac{\alpha}{2\pi f_A}
\sim 10^{-12}\ {\rm GeV}^{-1}          \left(
         \frac{10^{9}\, {\rm GeV}}{f_A}\right)
,
\end{equation}
which allows for various experimental probes. 

\item The {\it majoron} -- the Nambu-Goldstone boson arising from the breaking of global lepton number 
symmetry\cite{Chikashige:1980ui,Gelmini:1980re}  at a high energy scale $f_L=v_L$  -- thereby explaining the smallness of the masses of the left-handed SM active neutrinos by a see saw relation involving the electroweak scale Dirac-type mass $M_D = F\,v$ and the large Majorana-type mass $M_M=y\,f_L$ of extra right-handed SM singlet neutrinos,
\begin{equation}
\label{seesaw}
m_{\nu} = - M_D M_M^{-1} M_D^T = -  F\,y^{-1}\,F^T\ \frac{v ^2}{f_{L}}
= 0.6\,{\rm eV}  \left( \frac{10^{12}\,{\rm GeV}}{f_L} \right)
\left( \frac{-  F\,y^{-1}\,F^T}{10^{-2}}\right)
\,.
\end{equation}               

\item {\it Familons} arising from the breaking of global family symmetries\cite{Wilczek:1982rv,Berezhiani:1990wn,Jaeckel:2013uva}.                    

\item {\it Closed string axions\cite{Witten:1984dg,Conlon:2006tq,Cicoli:2012sz}} -- Kaluza-Klein zero modes of antisymmetric tensor fields -- the latter belonging to the massless spectrum of the
bosonic string propagating in ten dimensions. 
Their number $n_{\rm ax}$ is determined by the topology of the compactified manifold.

\item {\it Accidental axion and ALPs} arising from the breaking of accidental global $U(1)$ symmetries that appear as
low energy remnants of exact discrete symmetries -- the latter being postulated in purely field theoretic set ups\cite{Georgi:1981pu,Dias:2014osa} 
or occuring automatically in orbifold 
compactifications of the heterotic string\cite{Choi:2009jt}.

\end{itemize}
                                          
Therefore, searches for the axion $A$ -- the linear combination coupling to the topological charge density in QCD, Eq. (\ref{axion_def}) -- and further ALPs $a_j$  -- the  $n_{\rm ax-1}$ Nambu-Goldstone bosons perpendicular to the axion  in field space  -- are theoretically very well motivated. Moreover, as we will review next, their existence is  also suggested on cosmological and astrophysical grounds.

\section{\label{sec:hints}Physics Case for Axions and ALPs}

\subsection{Cold Dark Matter}

Axions and/or ALPs -- if they are pseudo Nambu-Goldstone bosons, i.e. if they have a (small) mass due to non-perturbative effects or explicit symmetry breaking -- are excellent cold dark matter candidates. In fact, for large symmetry breaking scales, axions and ALPs are very 
long lived. They are produced in the early universe via the vacuum realignment mechanism as a coherent state of many, extremely non-relativistic particles in the form of a classical, spatially coherent oscillating field\cite{Preskill:1982cy,Abbott:1982af,Dine:1982ah}.  
Neglecting anharmonic effects, today's (time $t_0$)  fraction of axion or ALP dark matter produced via the vacuum realignment mechanism is proportional to the spatially averaged field amplitude squared, $\langle a^2\rangle \equiv f_a^2 \langle \theta_a^2\rangle$,  at the time when the oscillations started, 
$t_{\rm osc}\simeq (3/2) \,m_a^{-1}(t_{\rm osc})$,\cite{Arias:2012az}
\begin{equation} 
\label{eq:CCDM}
R_a = \frac{\rho_a}{\rho_{\rm DM}}(t_0)\simeq 0.2\,  \sqrt{\frac{m_a (t_0)}{{\rm eV}}} 
\sqrt{\frac{m_a (t_0)}{m_a(t_{\rm osc})}}
\left(\frac{f_a}{10^{11}\, {\rm GeV}}\right)^2  \langle \theta_a^2\rangle\ 
 .
\end{equation}
Here, the indicated time-dependence of the mass arises from its temperature dependence, $m_a (t)\equiv m_a (T(t))$, taking into account possible 
plasma effects. 

The values of $\langle \theta_a^2\rangle$ depend crucially on whether the global symmetry breaking occured before inflation
ends and there was no symmetry restoration after inflation (``pre-inflationary SSB"), or the opposite (``post-inflationary SSB"), 
\begin{equation}
\langle\theta_a^2\rangle=
\left\{\begin{array}{ll}
\theta_i^2 + \left(\frac{H_I}{2\pi f_a}\right)^2,         &\text{if $f_a > \max \left( \frac{H_I}{2\pi} , \epsilon_{\rm eff} E_I \right)$,}\\[1ex]
\frac{\pi^2}{3},                                             &\text{if $f_a < \max \left( \frac{H_I}{2\pi} , \epsilon_{\rm eff}E_I \right)$.}
\end{array}
\right.
\end{equation}
Here, $H_I$ is the Hubble expansion rate during inflation, $E_I=\sqrt{\sqrt{\frac{3}{8\pi}}M_{\rm Pl}\, H_I}$ is the energy scale of inflation, and 
$\epsilon_{\rm eff}\in (0,1)$ is a reheating efficiency parameter. 

\begin{figure}[h]
\centerline{\includegraphics[width=0.6\linewidth]{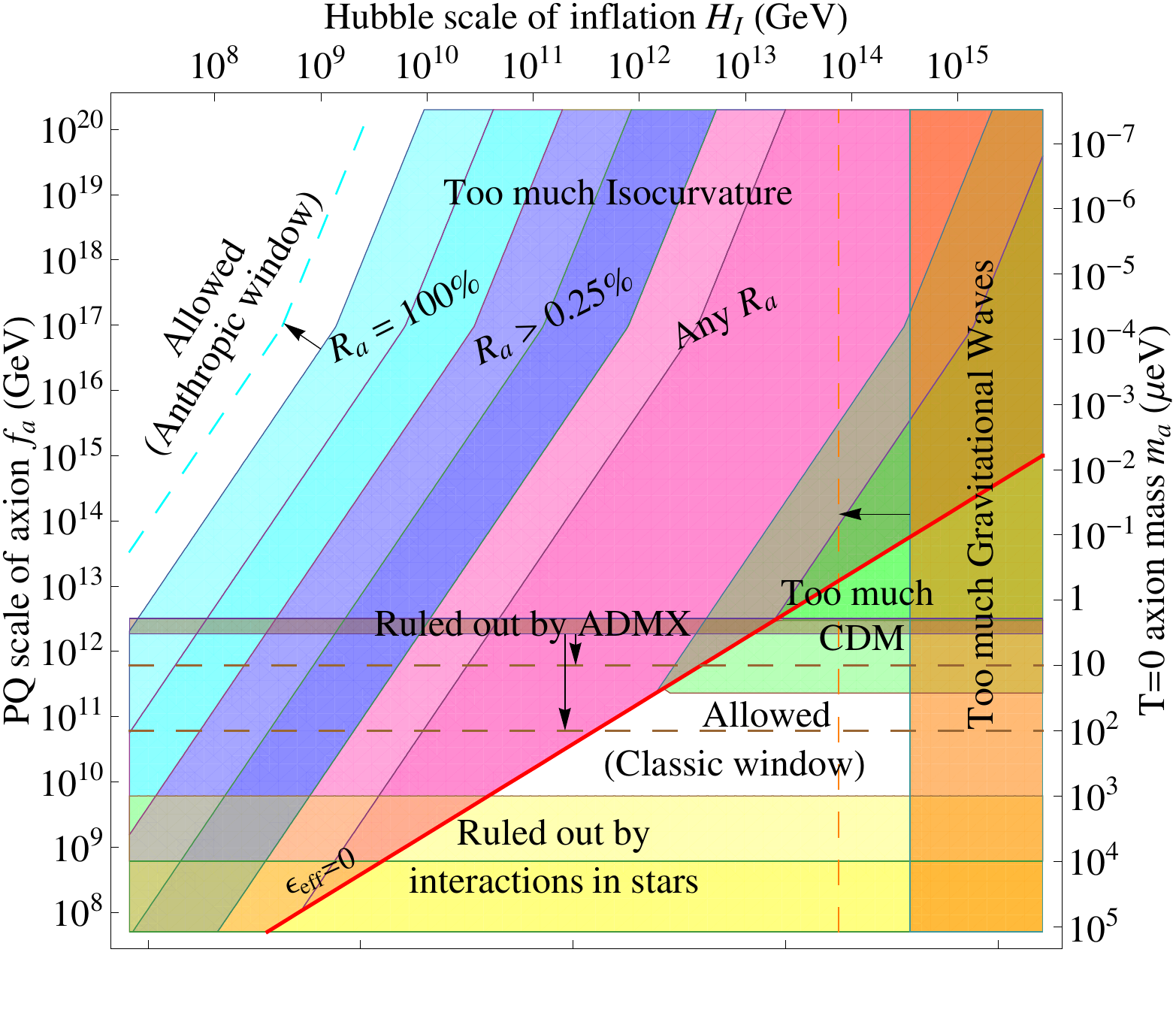}}
\caption[]{Disfavored (colored) and favored (white) regions of axion -- inflation parameter space\cite{Hertzberg:2008wr}.}
\label{fig:fAHI}
\end{figure}

In the first, ``pre-inflationary SSB" case, spatial variations in $\theta_a \equiv a/f_a$ are smoothed out over our Hubble volume. Then $\theta_i \equiv \langle \theta_a\rangle$ is an essentially environmental parameter drawn from a uniform distribution, $\theta_i \in [0,\pi]$, with a small variance, 
$\sigma_{\theta_a} = H_I/(2\pi f_a)$,
arising from quantum fluctuations during inflation. These fluctuations provide a lower bound on $R_a$. The non-observation 
of the associated isocurvature fluctuations in the anisotropies of the temperature of the cosmic microwave background (CMB) severely constrain axions and ALPs, disfavoring a sizeable region in $f_a$-$H_I$ parameter space\cite{Turner:1990uz,Fox:2004kb,Beltran:2006sq,Hertzberg:2008wr,Wantz:2009it}, as is exemplified in Fig. \ref{fig:fAHI}.  In fact, an $H_I$ of order $10^{14}\ {\rm GeV}$, as indicated by the recent detection of B-mode polarization in the CMB by BICEP2\cite{Ade:2014xna}
-- if the latter is interpreted as originating from primordial gravitational waves rather than from foreground dust\cite{Mortonson:2014bja,Flauger:2014qra} -- would strongly disfavor pre-inflationary SSB scenarios\cite{Marsh:2014qoa,Visinelli:2014twa} (possible ways out of this conclusion have been put forward in 
Refs.\cite{Dine:2004cq,Folkerts:2013tua,Higaki:2014ooa,Kawasaki:2014una}). 

This would favor the second, ``post-inflationary SSB" case,  in which the initial misalignment angle takes on different values in small (linear dimension $\sim m_a^{-1}(t_{\rm osc})$) patches in the universe, resulting in a  vanishing mean value, $\langle \theta_a\rangle =0$, and a variance $\sigma_{\theta_a} = \pi/\sqrt{3}$, due to small-scale variations. These variations again provide a lower bound on $R_a$ and, correspondingly, 
an upper bound on $f_a$. For the axion, this bound is of order\cite{Wantz:2009it,Bae:2008ue,DiValentino:2014zna} $f_A\lesssim {\rm few}\times 10^{11}\ {\rm GeV}$, see Fig. \ref{fig:fAHI}.  
Moreover, an additional cold dark matter population arises in the post-inflationary SSB scenario from the decay of topological defects, 
which radiates non-relativistic axions or ALPs. This additional contribution narrows down the parameter space of 
axions or ALPs even more, however with substantial uncertainties\cite{Wantz:2009it,Sikivie:2006ni,Hiramatsu:2012gg,Hiramatsu:2012sc}. 

In R-parity conserving supersymmetric models, more possibilities arise: 
cold dark matter might be a mixture of axions along with the lightest
SUSY particle (LSP)\cite{Baer:2013wza,Bae:2013hma}. Candidates for the LSP include then the lightest
neutralino, the gravitino, the axino, or a sneutrino.
In the case of a neutralino LSP, saxion and axino production 
in the early universe have a strong impact on 
the neutralino and axion abundance. 
For large values of $f_A$, saxions from the vacuum re-alignment mechanism may produce large relic dilution via entropy dumping. 

Recently, two groups reported the observation of an unidentified 3.55 keV line from galaxy clusters and from the Andromeda galaxy\cite{Bulbul:2014sua,Boyarsky:2014jta}. 
It is tempting to identify this line with the expected signal from two photon decay
of 7.1 keV mass ALP dark matter\cite{Higaki:2014zua,Jaeckel:2014qea}. To match the observed X-ray flux, but allowing for the likely possibility, that the ALP dark matter makes only a fraction $R_a \equiv \rho_a/\rho_{\rm DM}$ of the total density of dark matter, the required lifetime and thus decay constant of the ALP is (cf. Fig. \ref{fig:ALPplot}) 
\begin{equation}
  g_{a\gamma}  \sim 
R_a^{-1/2} 10^{-(17\div 18)}\,{\rm GeV}^{-1} \,; \hspace{6ex}
f_a \sim  C_{a\gamma} R_a^{1/2} 10^{14\div 15}\,{\rm GeV}\,.
\end{equation}
However, some observations such as the anomalous line strength in the Perseus cluster and the enhanced strength of the line emission in the cool cores of the Perseus, Ophiuchus and Centaurus clusters seem to be better fitted in models in which the unidentified 
line arises from the decay of  a 7.1 keV (scalar or fermionic) dark matter species into a very light ($m_a\lesssim 10^{-10}\,{\rm eV}$) ALP, that subsequently
converts to photons in astrophysical magnetic fields\cite{Cicoli:2014bfa,Conlon:2014wna}.

\subsection{Gamma Transparency of the Universe}

Gamma-ray spectra from distant active galactic nuclei (AGN) should show
an energy and redshift-dependent exponential attenuation, $\exp (-\tau (E,z))$,
due to $e^+ e^-$ pair production off the extragalactic background light (EBL) -- the stellar and
dust-reprocessed light accumulated during the cosmological evolution following
the era of re-ionization. In fact, Fermi\cite{Ackermann:2012sza} and
H.E.S.S.\cite{Abramowski:2012ry} have put sensible constraints on the EBL using their recent first 
detection of this effect. However, there is growing 
evidence\cite{Aharonian:2007wc,De Angelis:2007dy,Horns:2012fx,Rubtsov:2014uga} for an anomalous transparency of
the universe for gamma-rays at large optical depth, $\tau\gtrsim 2$.
This may be explained by photon $\leftrightarrow$ ALP oscillations:
the conversion of gamma rays into ALPs in the
magnetic fields around AGNs or in the intergalactic medium, followed by their unimpeded
travel towards our galaxy and the consequent reconversion into photons in the galactic/intergalactic magnetic
fields\cite{De Angelis:2007dy,Simet:2007sa,SanchezConde:2009wu,Meyer:2013pny,Tavecchio:2014yoa}. This explanation requires a very light ALP, which couples to two photons with strength\cite{Meyer:2013pny} (cf. Fig. \ref{fig:ALPplot}),
\begin{equation}
 g_{a\gamma}\gtrsim  10^{-12} \ {\rm GeV}^{-1}
; \hspace{6ex}
 m_{a}\lesssim 10^{-7}\  {\rm eV}.
\label{dec_const_transp}
\end{equation}

\subsection{Cosmic ALP Background Radiation}

\begin{figure}
\centerline{\includegraphics[width=0.5\linewidth]{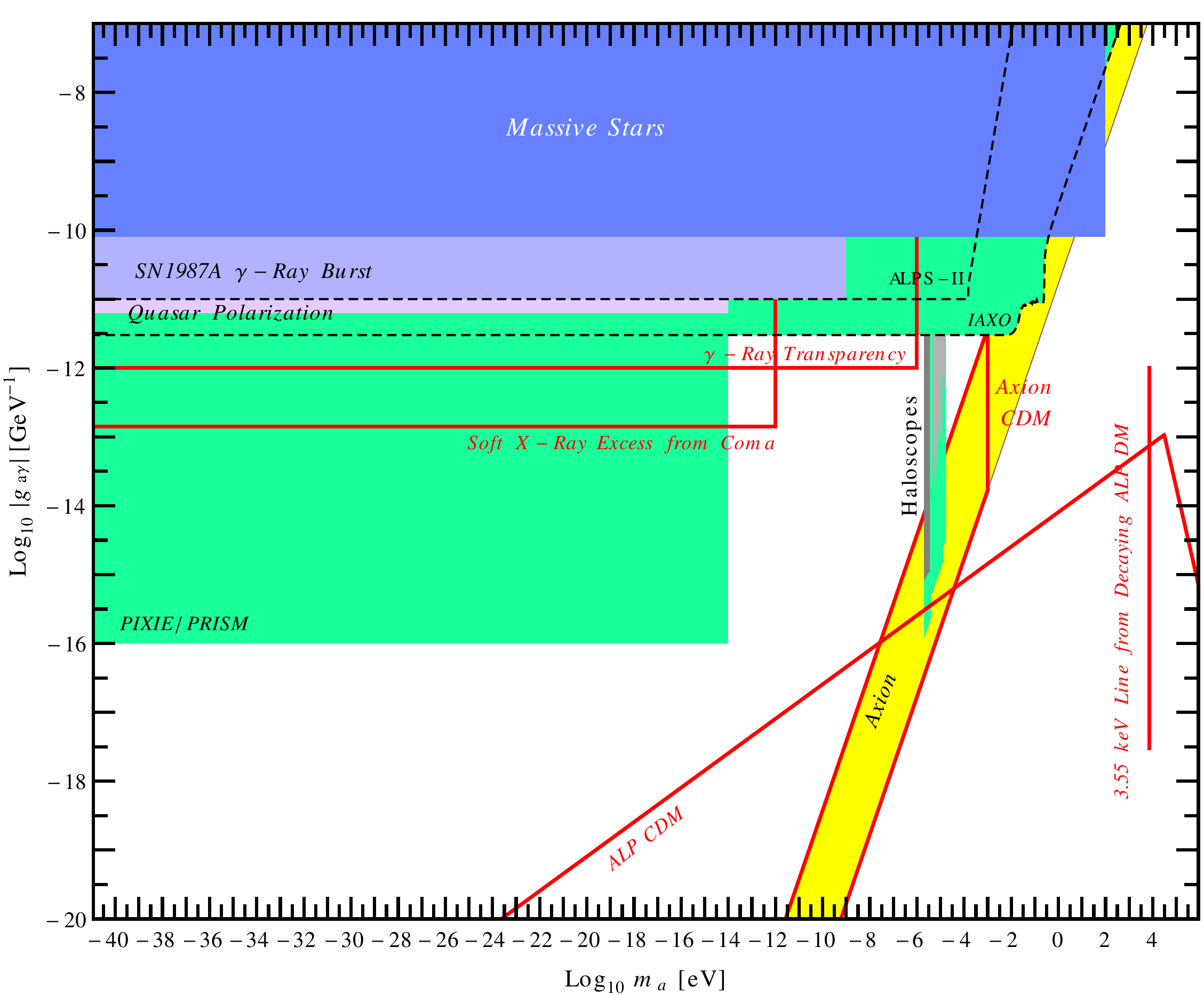}}
\caption[]{Axion and ALP photon coupling versus mass\cite{Dias:2014osa}. The figure shows the prediction for the axion (yellow band) and excluded regions arising from the non-observation of an anomalous energy loss of massive stars due to axion or ALP emission\cite{Friedland:2012hj}, of a $\gamma$-ray burst from SN 1987A
due to conversion of an initial ALP burst in the galactic magnetic field\cite{Brockway:1996yr,Grifols:1996id},  of  changes in quasar polarizations due to photon-ALP oscillations\cite{Payez:2012vf}, and of dark matter axions or ALPs converted into photons in microwave cavities placed in magnetic fields.
Axions and ALPs with parameters in the regions surrounded by the red lines may
constitute all or a part of cold dark matter (CDM), explain the cosmic $\gamma$-ray transparency, and the soft X-ray excess from Coma.
The green regions are the projected sensitivities of the light-shining-through-wall experiment ALPS II, of the helioscope IAXO, of the
haloscopes ADMX and ADMX-HF, and of the PIXIE or PRISM CMB observatories.}
\label{fig:ALPplot}
\end{figure}

There are observational hints on extra dark radiation in the primordial plasma during big bang nucleosynthesis and before photon decoupling -- beyond the one from the three active neutrino species -- at the one to three sigma level\cite{Ade:2013zuv,Cooke:2013cba}. 
Intriguingly, a cosmic ALP background (CAB) radiation corresponding to an effective number $\triangle N_{\rm eff}\sim 0.5$ of extra neutrinos species 
can be naturally produced by the decay of a heavy ($\sim 10^6\,{\rm GeV}$) modulus with Planck mass suppressed couplings\cite{Cicoli:2012aq,Higaki:2012ar}.  
In fact, an observed soft X-ray excess in the Coma cluster may be explained by the conversion of such a CAB 
into photons in the cluster magnetic field\cite{Conlon:2013txa,Angus:2013sua}.
This explanation requires
that the CAB spectrum is peaked in the soft X-ray region and that the ALP coupling and mass satisfy
\begin{equation}
g_{a\gamma}\gtrsim  10^{-13} \ {\rm GeV}^{-1}\, \sqrt{0.5/\triangle N_{\rm eff}}\,; \hspace{6ex} m_a\lesssim 10^{-12}\ {\rm eV},
\end{equation}
respectively,
overlapping with the
parameter range \eqref{dec_const_transp} preferred by the ALP solution of the gamma-ray transparency puzzle, as is apparent in
Fig. \ref{fig:ALPplot}.

\section{\label{sec:experiments}Axion and ALP Experiments and Observatories}

We have seen in the last section that there are strong theoretical, cosmological and astrophysical hints 
suggesting the existence of the axion plus additional  two to three ALPs (cf. red regions in Fig. \ref{fig:ALPplot}). 
Fortunately, a sizeable part of the favored regions in axion and ALP parameter space can be explored in the 
foreseeable future with experimental searches based on  axion or ALP photon oscillations in magnetic fields
(cf. green regions in Fig. \ref{fig:ALPplot}), 
as we review in this section.

\subsection{Haloscope Searches}

Haloscopes directly search for  galactic halo dark matter axions and ALPs in the laboratory. 
Currently, the most sensitive ones exploit electromagnetic cavites placed in  
a strong magnet\cite{Sikivie:1983ip}. They aim for the detection of electromagnetic power arising from
the conversion of dark matter axions or ALPs into real photons, with frequency 
$
\nu=m_{a}/(2\pi )=0.24\ {\rm GHz}\times (m_{a}/\mu{\rm eV})$.
The best sensitivity is reached on resonance, the power output then being proportional to the quality factor of the cavity. 
The Axion Dark Matter eXperiment (ADMX) has indeed reached recently the sensitivity to probe 
axion dark matter\cite{Asztalos:2009yp} (see Fig. \ref{fig:ALPplot}).
The ongoing experiments ADMX-II and ADMX-High Frequency (HF) aim to explore the green regions labelled ``Haloscopes" in the same figure. Further haloscope opportunities
in complementary mass ranges 
may arise from recycling available microwave cavities and magnets at accelerator laboratories\cite{Baker:2011na,Horns:2013ira}. 
 
Other new haloscope concepts are also being investigated.  
A microwave Fabry-Perot resonator in a spatially varying magnetic field may be exploited\cite{Rybka:2014cya} to search for axion/ALP dark matter
with masses above 40 $\mu$eV. 
Converted photons from axion/ALP dark matter could be focused in a manner similar to a dish antenna, allowing for broad-band 
searches\cite{Horns:2012jf}. 
Precision magnetometry may be exploited to search for oscillating nuclear electric dipole moments induced by the oscillating galactic
dark matter axion field\cite{Budker:2013hfa}. 
DM axions/ALPs cause an oscillating electric current
to flow along magnetic field lines. The corresponding oscillating magnetic field may be amplified using 
an LC circuit and then detected by precision magnetometry\cite{Sikivie:2013laa}. 

\subsection{Light-Shining-Through-Wall Searches}

Light-Shining-Through-Wall (LSW) experiments aim both for production and detection of axions and ALPs in the laboratory. 
This is done by sending laser photons along a strong magnetic field, allowing for their conversion into axions or ALPs, towards a blocking wall, behind of which the latter may then reconvert, 
again in a strong magnetic field, into photons, the latter being susceptible to detection\cite{Anselm:1986gz,VanBibber:1987rq,Redondo:2010dp}. 
The Any Light Particle Search (ALPS I) 
experiment at DESY has currently established the best sensitivity of LSW experiments\cite{Ehret:2010mh}. 
Its successor experiment ALPS II\cite{Bahre:2013ywa} proposes to use 10+10 straightened 
HERA magnets\cite{Ringwald:2003nsa}, a high-power laser
system, a superconducting low-background detector and the pioneering realization of an optical
regeneration cavity\cite{Hoogeveen:1990vq,Sikivie:2007qm}. 
It aims 
to tackle some of the ALP 
parameter space favored by astrophysical observations, cf. the light-green region in 
Fig.~\ref{fig:ALPplot} labelled by ``ALPS II". 
LSW experiments in other spectral ranges, notably in the 
microwave\cite{Hoogeveen:1992nq,Caspers:2009cj} 
and in the X-ray ranges\cite{Rabadan:2005dm,Dias:2009ph}, 
are still in the pioneering stage\cite{Battesti:2010dm,Betz:2013dza} and do not seem to be competitive in the foreseeable future.

\subsection{Helioscope Searches}

Helioscopes aim to detect solar axions and ALPs produced by their conversion into photons inside of a strong magnet pointing towards the Sun\cite{Sikivie:1983ip}. The CERN Axion Solar Telescope (CAST), employing an LHC dipole test magnet,  sets currently the best helioscope 
limit\cite{Andriamonje:2007ew,Arik:2013nya}.
A proposed next-generation axion helioscope, dubbed the International Axion Observatory (IAXO),
envisions a dedicated superconducting toroidal magnet with much bigger aperture than CAST,
a detection system consisting of large X-ray telescopes coupled to
ultra-low background X-ray detectors, and a large, robust tracking system\cite{Armengaud:2014gea}, and aims at the 
sensitivity shown in Fig.~\ref{fig:ALPplot}. 

\section{Summary}

There is a strong physics case for the axion and ALPs.  They occur naturally 
in many 
theoretically appealing ultraviolet completions of the SM. 
They are dark matter
candidates and can explain the anomalous cosmic gamma ray transparency, 
soft X-ray excesses from galaxy clusters, and the unidentified 3.55 keV line from Andromeda and galaxy clusters. 
A significant portion of their parameter space will be tackeled in this 
decade by experiments. 
Stay tuned!


\section*{References}

\begin{thebibliography}{99}

\bibitem{Peccei:1977hh}
  R.~D.~Peccei and H.~R.~Quinn,
  Phys.\ Rev.\ Lett.\  {\bf 38} (1977) 1440.

\bibitem{Weinberg:1977ma}
  S.~Weinberg,
  Phys.\ Rev.\ Lett.\  {\bf 40} (1978) 223.

\bibitem{Wilczek:1977pj}
  F.~Wilczek,
  Phys.\ Rev.\ Lett.\  {\bf 40} (1978) 279.

\bibitem{Chikashige:1980ui}
  Y.~Chikashige, R.~N.~Mohapatra and R.~D.~Peccei,
  Phys.\ Lett.\ B {\bf 98} (1981) 265.

\bibitem{Gelmini:1980re}
  G.~B.~Gelmini and M.~Roncadelli,
  Phys.\ Lett.\ B {\bf 99} (1981) 411.

\bibitem{Wilczek:1982rv}
  F.~Wilczek,
  Phys.\ Rev.\ Lett.\  {\bf 49} (1982) 1549.

\bibitem{Berezhiani:1990wn}
  Z.~G.~Berezhiani and M.~Y.~Khlopov,
  Sov.\ J.\ Nucl.\ Phys.\  {\bf 51} (1990) 739.

\bibitem{Jaeckel:2013uva}
  J.~Jaeckel,
  Phys.\ Lett.\ B {\bf 732} (2014) 1
  [arXiv:1311.0880 [hep-ph]].

\bibitem{Witten:1984dg}
  E.~Witten,
  Phys.\ Lett.\ B {\bf 149} (1984) 351.

\bibitem{Conlon:2006tq}
  J.~P.~Conlon,
  JHEP {\bf 0605} (2006) 078
  [hep-th/0602233].

\bibitem{Cicoli:2012sz}
  M.~Cicoli, M.~Goodsell and A.~Ringwald,
  JHEP {\bf 1210} (2012) 146.

\bibitem{Georgi:1981pu}
  H.~M.~Georgi, L.~J.~Hall and M.~B.~Wise,
  Nucl.\ Phys.\ B {\bf 192} (1981) 409.

\bibitem{Dias:2014osa}
  A.~G.~Dias {\em et al.}, 
  JHEP {\bf 1406} (2014) 037.

\bibitem{Choi:2009jt}
  K.~-S.~Choi {\em et al.},  
  Phys.\ Lett.\ B {\bf 675} (2009) 381
  [arXiv:0902.3070 [hep-th]].

\bibitem{Preskill:1982cy}
  J.~Preskill, M.~B.~Wise and F.~Wilczek,
  Phys.\ Lett.\ B {\bf 120} (1983) 127.

\bibitem{Abbott:1982af}
  L.~F.~Abbott and P.~Sikivie,
  Phys.\ Lett.\ B {\bf 120} (1983) 133.

\bibitem{Dine:1982ah}
  M.~Dine and W.~Fischler,
  Phys.\ Lett.\ B {\bf 120} (1983) 137.

\bibitem{Arias:2012az}
  P.~Arias {\em et al.},  
  JCAP {\bf 1206} (2012) 013
  [arXiv:1201.5902 [hep-ph]].

\bibitem{Turner:1990uz}
  M.~S.~Turner and F.~Wilczek,
  Phys.\ Rev.\ Lett.\  {\bf 66} (1991) 5.

\bibitem{Fox:2004kb}
  P.~Fox, A.~Pierce and S.~D.~Thomas,
  hep-th/0409059.

\bibitem{Beltran:2006sq}
  M.~Beltran, J.~Garcia-Bellido and J.~Lesgourgues,
  Phys.\ Rev.\ D {\bf 75} (2007) 103507.

\bibitem{Hertzberg:2008wr}
  M.~P. Hertzberg, M.~Tegmark and F.~Wilczek,
  Phys.\ Rev.\ D {\bf 78} (2008) 083507.

\bibitem{Wantz:2009it}
  O.~Wantz and E.~P.~S.~Shellard,
  Phys.\ Rev.\ D {\bf 82} (2010) 123508.

\bibitem{Ade:2014xna}
  P.~A.~R.~Ade {\it et al.}  [BICEP2 Collaboration],
  Phys.\ Rev.\ Lett.\  {\bf 112} (2014) 241101.

\bibitem{Mortonson:2014bja}
  M.~J.~Mortonson and U.~Seljak,
  arXiv:1405.5857 [astro-ph.CO].

\bibitem{Flauger:2014qra}
  R.~Flauger, J.~C.~Hill and D.~N.~Spergel,
  arXiv:1405.7351 [astro-ph.CO].

\bibitem{Marsh:2014qoa}
  D.~J.~E.~Marsh, D.~Grin, R.~Hlozek and P.~G.~Ferreira,
  arXiv:1403.4216 [astro-ph.CO].

\bibitem{Visinelli:2014twa}
  L.~Visinelli and P.~Gondolo,
  arXiv:1403.4594 [hep-ph].

\bibitem{Dine:2004cq}
  M.~Dine and A.~Anisimov,
  JCAP {\bf 0507} (2005) 009
  [hep-ph/0405256].

\bibitem{Folkerts:2013tua}
  S.~Folkerts, C.~Germani and J.~Redondo,
  Phys.\ Lett.\ B {\bf 728} (2014) 532.

\bibitem{Higaki:2014ooa}
  T.~Higaki, K.~S.~Jeong and F.~Takahashi,
  Phys.\ Lett.\ B {\bf 734} (2014) 21.

\bibitem{Kawasaki:2014una}
  M.~Kawasaki, N.~Kitajima and F.~Takahashi,
  arXiv:1406.0660 [hep-ph].

\bibitem{Bae:2008ue}
  K.~J.~Bae, J.~-H.~Huh and J.~E.~Kim,
  JCAP {\bf 0809} (2008) 005
  [arXiv:0806.0497 [hep-ph]].

\bibitem{DiValentino:2014zna}
  E.~Di Valentino {\it et al.},  
  arXiv:1405.1860 [astro-ph.CO].

\bibitem{Sikivie:2006ni}
  P.~Sikivie,
  Lect.\ Notes Phys.\  {\bf 741} (2008) 19
  [astro-ph/0610440].

\bibitem{Hiramatsu:2012gg}
  T.~Hiramatsu {\it et al.},  
  Phys.\ Rev.\ D {\bf 85} (2012) 105020
   [Erratum-ibid.\ D {\bf 86} (2012) 089902].

\bibitem{Hiramatsu:2012sc}
  T.~Hiramatsu {\it et al.},  
  JCAP {\bf 1301} (2013) 001
  [arXiv:1207.3166 [hep-ph]].

\bibitem{Baer:2013wza}
  H.~Baer,
  arXiv:1310.1859 [hep-ph].

\bibitem{Bae:2013hma}
  K.~J.~Bae, H.~Baer and E.~J.~Chun,
  JCAP {\bf 1312} (2013) 028
  [arXiv:1309.5365 [hep-ph]].

\bibitem{Bulbul:2014sua}
  E.~Bulbul {\it et al.}, 
  Astrophys.\ J.\  {\bf 789} (2014) 13
  [arXiv:1402.2301 [astro-ph.CO]].

\bibitem{Boyarsky:2014jta}
  A.~Boyarsky, O.~Ruchayskiy, D.~Iakubovskyi and J.~Franse,
  arXiv:1402.4119 [astro-ph.CO].

\bibitem{Higaki:2014zua}
  T.~Higaki, K.~S.~Jeong and F.~Takahashi,
  Phys.\ Lett.\ B {\bf 733} (2014) 25.

\bibitem{Jaeckel:2014qea}
  J.~Jaeckel, J.~Redondo and A.~Ringwald,
  Phys.\ Rev.\ D {\bf 89} (2014) 103511.

\bibitem{Friedland:2012hj}
  A.~Friedland, M.~Giannotti and M.~Wise,
  Phys.\ Rev.\ Lett.\  {\bf 110} (2013) 061101.

\bibitem{Brockway:1996yr}
  J.~W.~Brockway, E.~D.~Carlson and G.~G.~Raffelt,
  Phys.\ Lett.\ B {\bf 383} (1996) 439.

\bibitem{Grifols:1996id}
  J.~A.~Grifols, E.~Masso and R.~Toldra,
  Phys.\ Rev.\ Lett.\  {\bf 77} (1996) 2372
  [astro-ph/9606028].

\bibitem{Payez:2012vf}
  A.~Payez, J.~R.~Cudell and D.~Hutsemekers,
  JCAP {\bf 1207} (2012) 041.

\bibitem{Cicoli:2014bfa}
  M.~Cicoli, J.~P.~Conlon, M.~C.~D.~Marsh and M.~Rummel,
  arXiv:1403.2370 [hep-ph].

\bibitem{Conlon:2014wna}
  J.~P.~Conlon and A.~J.~Powell,
  arXiv:1406.5518 [hep-ph].

\bibitem{Ackermann:2012sza}
  M.~Ackermann {\it et al.}  [Fermi-LAT Collaboration],
  Science {\bf 338} (2012) 1190.

\bibitem{Abramowski:2012ry}
  A.~Abramowski {\it et al.}  [H.E.S.S. Collaboration],
  arXiv:1212.3409 [astro-ph.HE].

\bibitem{Aharonian:2007wc}
  F.~Aharonian {\it et al.}  [H.E.S.S. Collaboration],
  Astron.\ Astrophys.\  {\bf 475} (2007) L9.

\bibitem{De Angelis:2007dy}
  A.~De Angelis, M.~Roncadelli and O.~Mansutti,
  Phys.\ Rev.\ D {\bf 76} (2007) 121301.

\bibitem{Horns:2012fx}
  D.~Horns and M.~Meyer,
  JCAP {\bf 1202} (2012) 033
  [arXiv:1201.4711 [astro-ph.CO]].

\bibitem{Rubtsov:2014uga}
  G.~I.~Rubtsov and S.~V.~Troitsky,
  arXiv:1406.0239 [astro-ph.HE].

\bibitem{Simet:2007sa}
  M.~Simet, D.~Hooper and P.~D.~Serpico,
  Phys.\ Rev.\ D {\bf 77} (2008) 063001.

\bibitem{SanchezConde:2009wu}
  M.~A.~Sanchez-Conde {\it et al.},  
  Phys.\ Rev.\ D {\bf 79} (2009) 123511.

\bibitem{Meyer:2013pny}
  M.~Meyer, D.~Horns and M.~Raue,
  Phys.\ Rev.\ D {\bf 87} (2013) 035027.

\bibitem{Tavecchio:2014yoa}
  F.~Tavecchio, M.~Roncadelli and G.~Galanti,
  arXiv:1406.2303 [astro-ph.HE].

\bibitem{Ade:2013zuv}
  P.~A.~R.~Ade {\it et al.}  [Planck Collaboration],
  arXiv:1303.5076 [astro-ph.CO].

\bibitem{Cooke:2013cba}
  R.~Cooke {\it et al.},  
  arXiv:1308.3240 [astro-ph.CO].

\bibitem{Cicoli:2012aq}
  M.~Cicoli, J.~P.~Conlon and F.~Quevedo,
  Phys.\ Rev.\ D {\bf 87} (2013) 043520.

\bibitem{Higaki:2012ar}
  T.~Higaki and F.~Takahashi,
  JHEP {\bf 1211} (2012) 125
  [arXiv:1208.3563 [hep-ph]].

\bibitem{Conlon:2013txa}
  J.~P.~Conlon and M.~C.~D.~Marsh,
  Phys.\ Rev.\ Lett.\  {\bf 111} (2013) 15,  151301.

\bibitem{Angus:2013sua}
  S.~Angus {\it et al.},  
  arXiv:1312.3947 [astro-ph.HE].

\bibitem{Sikivie:1983ip}
  P.~Sikivie,
  Phys.\ Rev.\ Lett.\  {\bf 51} (1983) 1415
   [Erratum-ibid.\  {\bf 52} (1984) 695].

\bibitem{Asztalos:2009yp}
  S.~J.~Asztalos {\it et al.}  [ADMX Collaboration],
  Phys.\ Rev.\ Lett.\  {\bf 104} (2010) 041301.

\bibitem{Baker:2011na}
  O.~K.~Baker {\it et al.},  
  Phys.\ Rev.\ D {\bf 85} (2012) 035018
  [arXiv:1110.2180 [physics.ins-det]].

\bibitem{Horns:2013ira}
  D.~Horns, A.~Lindner, A.~Lobanov and A.~Ringwald,
  arXiv:1309.4170 [physics.ins-det].

\bibitem{Rybka:2014cya}
  G.~Rybka and A.~Wagner,
  arXiv:1403.3121 [physics.ins-det].

\bibitem{Horns:2012jf}
  D.~Horns {\it et al.},  
  JCAP {\bf 1304} (2013) 016
  [arXiv:1212.2970].

\bibitem{Budker:2013hfa}
  D.~Budker {\it et al.},  
  Phys.\ Rev.\ X {\bf 4} (2014) 021030
  [arXiv:1306.6089 [hep-ph]].

\bibitem{Sikivie:2013laa}
  P.~Sikivie, N.~Sullivan and D.~B.~Tanner,
  arXiv:1310.8545 [hep-ph].

\bibitem{Anselm:1986gz}
  A.~A.~Anselm,
  Yad.\ Fiz.\  {\bf 42} (1985) 1480.
  
\bibitem{VanBibber:1987rq}
  K.~Van Bibber {\it et al.},  
  Phys.\ Rev.\ Lett.\  {\bf 59} (1987) 759.
  
\bibitem{Redondo:2010dp}
  J.~Redondo and A.~Ringwald,
  Contemp.\ Phys.\  {\bf 52} (2011) 211
  [arXiv:1011.3741 [hep-ph]].
 
\bibitem{Ehret:2010mh}
  K.~Ehret {\it et al.} [ALPS Collaboration],  
  Phys.\ Lett.\ B {\bf 689} (2010) 149.

\bibitem{Bahre:2013ywa}
  R.~Baehre {\it et al.} [ALPS Collaboration],  
  JINST {\bf 8} (2013) T09001.

\bibitem{Ringwald:2003nsa}
  A.~Ringwald,
  Phys.\ Lett.\ B {\bf 569} (2003) 51
  [hep-ph/0306106].

\bibitem{Hoogeveen:1990vq}
  F.~Hoogeveen and T.~Ziegenhagen,
  Nucl.\ Phys.\ B {\bf 358} (1991) 3.

\bibitem{Sikivie:2007qm}
  P.~Sikivie, D.~B.~Tanner and K.~van Bibber,
  Phys.\ Rev.\ Lett.\  {\bf 98} (2007) 172002.

\bibitem{Hoogeveen:1992nq}
  F.~Hoogeveen,
  Phys.\ Lett.\ B {\bf 288} (1992) 195.

\bibitem{Caspers:2009cj}
  F.~Caspers, J.~Jaeckel and A.~Ringwald,
  JINST {\bf 4} (2009) P11013
  [arXiv:0908.0759 [hep-ex]].

\bibitem{Rabadan:2005dm}
  R.~Rabadan, A.~Ringwald and K.~Sigurdson,
  Phys.\ Rev.\ Lett.\  {\bf 96} (2006) 110407.

\bibitem{Dias:2009ph}
  A.~G.~Dias and G.~Lugones,
  Phys.\ Lett.\ B {\bf 673} (2009) 101
  [arXiv:0902.0749 [hep-ph]].

\bibitem{Battesti:2010dm}
  R.~Battesti {\it et al.},  
  Phys.\ Rev.\ Lett.\  {\bf 105} (2010) 250405
  [arXiv:1008.2672 [hep-ex]].

\bibitem{Betz:2013dza}
  M.~Betz {\it et al.},  
  Phys.\ Rev.\ D {\bf 88} (2013) 7,  075014 
  [arXiv:1310.8098 [physics.ins-det]].

\bibitem{Andriamonje:2007ew}
  S.~Andriamonje {\it et al.}  [CAST Collaboration],
  JCAP {\bf 0704} (2007) 010
  [hep-ex/0702006].

\bibitem{Arik:2013nya}
  M.~Arik {\it et al.}  [CAST Collaboration],  
  Phys.\ Rev.\ Lett.\  {\bf 112} (2014) 091302.

\bibitem{Armengaud:2014gea}
  E.~Armengaud {\it et al.}   [IAXO Collaboration],  
  JINST {\bf 9} (2014) T05002.

\end{thebibliography}
\end{document}